# Novel carbon materials: new tunneling systems

(**Review Article**)

M.A. Strzhemechny and A.V. Dolbin

*B. Verkin Institute for Low Temperature Physics and Engineering of the National Academy of Sciences of Ukraine*
*47 Lenin Ave., Kharkov 61103, Ukraine*
E-mail: strzhemechny@ilt.kharkov.ua

This review covers recent achievements in the studies of quantum properties of the novel carbon materials (fullerite $C_{60}$ and bundles of single-walled nanotubes (SWNT)) saturated with such light-mass species as helium isotopes, the homonuclear molecular hydrogens, and neon. It is shown that even some heavy dopants demonstrate kinetic phenomena, in which coherent effects play an essential role. Two theoretical concepts are surveyed which have been suggested for the explanation of the anomalous phenomena in saturation kinetics and linear thermal expansion of doped $C_{60}$. Most unusual effects have been also observed in the low-temperature radial expansion of bundles of single-walled carbon nanotubes saturated with the helium isotopes. First, it was shown that low-temperature radial expansion of pure SWNT is negative, i.e., a nanotube shrinks with warming. Second, saturation of SWNT bundles with the helium isotopes entails a huge increase of the negative expansion effect, when the dopant is He. So far, no detailed physical picture has been put forward. It is worth mentioning that the dynamics of a single helium atom on an isolated nanotube corresponds to that of a tight-bound quasiparticle with a band width of about 10 K.

## Contents

*To Prof. V.G. Manzhelli, a good collegua and a great teacher*.

Although the time elapsed since the discovery of the new carbon nanomaterials is rather short the Nobel Prize Committee had twice recognized [1,2] their striking novelty and importance for fundamental science. Researchers attacked these materials from every imaginable angle and, seemingly, after 20 years of these efforts really nothing new could be found in the properties of these materials. However, it turned out that the new is always at hand. This brief review presents unusual results recently obtained for $C_{60}$ and single-walled carbon nanotubes doped with light-mass species like helium and hydrogen isotopes. All experimental results have been obtained at the Verkin Institute for Low Temperature Physics and Engineering.

## 1. $C_{60}$ fullerite properties

For easier understanding we give below a short compendium of the properties of the $C_{60}$ which we will be mentioning in the review. At room temperature $C_{60}$ adopts a fcc structure [3], determined mostly by van der Waals forces. At room temperature the lattice parameter is 14.16 Å and the nearest neighbor distance is close to 10 Å; the molecules rotate more or less freely.

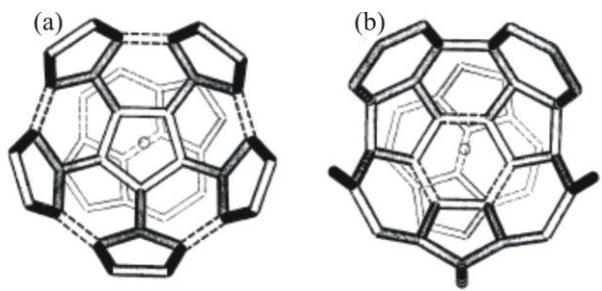

*Fig. 1.* A double bond (elevated electron density), marked with a tiny circle, of any molecule faces either a hexagon (b) (elevated electron density) or a pentagon (a) (lower electron density). The latter mutual orientation is energetically preferable.

Near 260 K, molecular rotations cease to be completely chaotic because the anisotropic interactions become comparable with the kinetic energy of rotation. Below 260 K a (partial) orientational ordering sets in, the molecules executing thermally activated jump-like rotations around certain axes [4]. As the temperature is decreased, the rotational jumps become progressively slower to stop completely at approximately 100 K. Below this point one deals with a new state which was termed "orientational glass". The reason for this is as follows. The fullerene molecule $C_{60}$ is highly symmetric. In particular, it has many 3-folds axes, which go through two opposite hexagons. The molecular array between 100 and 260 K is such [5] that one of the double bonds in one molecule faces either a pentagon or a hexagon of the respective neighbor molecule as shown in Fig. 1. That means that in the former case the double bond, an entity with an elevated electron density, is closermost to a pentagon, which consists only of single bounds, that is, an entity with a depleted electron density. Evidently, the latter case is preferable as far as the electrostatic energy is concerned. Thus, when a molecule rotates through $\pi/6$ it goes from one potential well to a shallower or deeper one.

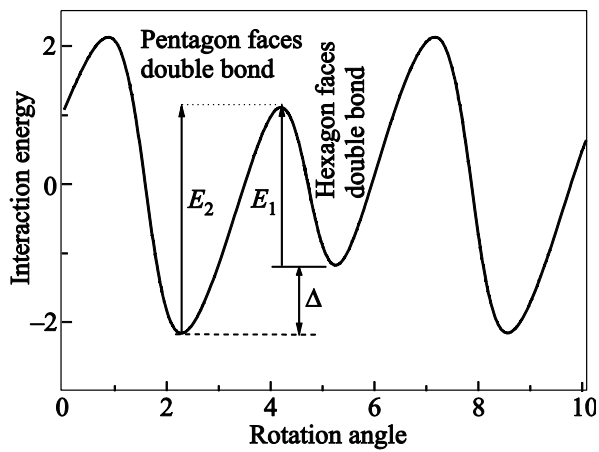

*Fig. 2.* A sketch of the electrostatic interaction energy of two neighboring $C_{60}$ molecule as a function of the rotation angle of one of these molecules.

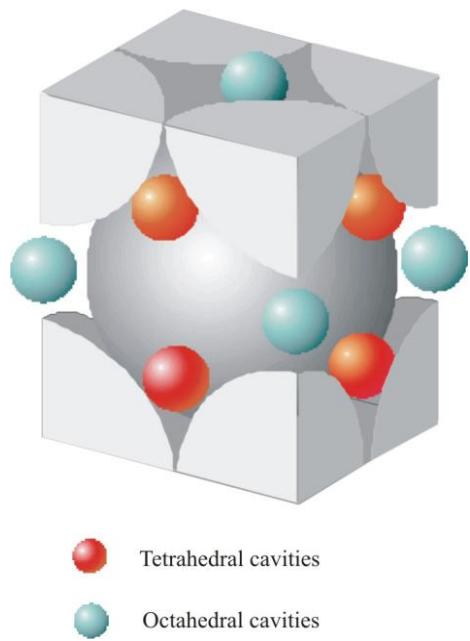

*Fig. 3.* Positions of the voids in the fullerite lattice.

The important moment is that the energy gain $\Delta$ in Fig. 2 by more than an order of magnitude exceeds the potential barrier (at least, $E_1$). The quantity $\Delta$ controls the thermodynamic probabilities while the barrier height determines the characteristic time $\tau$ of events. If $\tau$ becomes much longer than the typical experimental time, the corresponding events, even urged by the thermodynamics, could hardly happen. Because of that, as the temperature goes below approximately 100 K the molecules cease to rotate completely and the fullerite is said to be in the state of orientational glass. The fraction of energetically unfavorable mutual pairwise molecular orientations is frozen [6] at roughly 16%.

The large size of the molecules that constitute the lattice of the molecular crystal fullerite $C_{60}$ explains why the interstitial voids are sufficiently roomy to easily accommodate atoms and smaller molecules. The diameters of these voids (cf. Fig. 3) amount to 4.2 and 2.2 Å for octahedral and tetrahedral voids, respectively.

## 2. Experimental methods and samples

The experimental techniques employed in the studies under survey were high-precision dilatometry, high-precision low-pressure measurements, and powder x-ray diffraction. Every of these techniques necessitated rather involved special sample preparation procedures, which had first to be invented and then put into life. The high-precision dilatometry setup published in sufficient detail elsewhere [7,8] is unique as far as its accuracy of $2 \cdot 10^{-9}$ cm is concerned. Since the quantity measured is the uniaxial elongation caused by warming or cooling, the structure of the solid under study should be preferably cubic, which luckily is so with fullerite $C_{60}$. The best accuracy is reached for long enough samples (a few centimeters). No single crystal of $C_{60}$ of that size exists, therefore a powder $C_{60}$ was compressed under not too severe conditions (pressures below 1 MPa). Since $C_{60}$ saturated with various dopant species was to be studied, the samples were produced first by saturation (in Australia, gas pressure of about 200 MPa and temperature of 575 °C), then compressed (in Sweden) and shipped to Ukraine for final investigation.

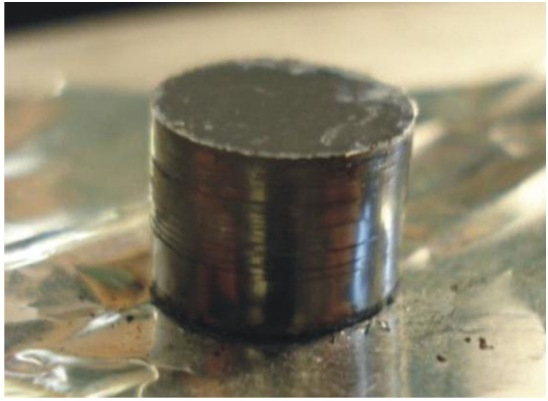

*Fig. 4.* A view of the sample of single-walled carbon nanotube bundles compacted so that axes of all nanotubes are perpendicular to the cylinder axis.

A still more stringent requirements apply to SWNT samples to be investigated using the dilatometric method. Usually the nanotube samples are bundles, in which the number and types of nanotubes may differ significantly. To tackle the task, nanotube samples consisting of unknown number of bundles were compressed to a cylinder (Fig. 4), in which all nanotubes, however randomly oriented, lay flat with each axis (albeit, arbitrarily twisted in plane) predominantly at the same particular distance from bottom. Such a configuration allowed measurement of the radial expansion coefficient [9]. Another technique that proved very efficient for measurements of saturation levels to high accuracy was high-precision low-pressure monitoring [10]. The x-ray diffraction studies were carried out in a standard way, the distinguishing feature being that saturation with light-mass dopants (He, $H_2$) was affected *in situ* at a constant low pressure (about 1 bar), which allowed accurate measurement of variations of the lattice parameter of $C_{60}$ in time and subsequent evaluation of the respective diffusion coefficients. Molecular hydrogen was also stuffed at elevated (temperatures up to 300 °C) and pressures (up to 30 kbar) in a special saturator.

### 3. Thermal expansion of pure and doped $C_{60}$ at low temperatures

Very first measurements [8] of nominally pure $C_{60}$ yielded negative linear expansion coefficients $\alpha$ below 5 K, which served as a big incentive to begin systematic studies as well as to improve accuracy. Since at low temperatures $C_{60}$ is in the orientational glass state, the first explanation [11] was based on the assumption that the states responsible for negative $\alpha$ are tunnel rotations of $C_{60}$ molecules.

More accurate later dilatometric investigations showed that pure $C_{60}$ does not feature negative expansivity. The peculiarity of low-temperature dilatometry measurements consists in the fact that only the initial warm-up run very often yields negative expansion coefficients. The next ones (after the first warm-up and cool-down runs) do not result in negative expansivities. But no matter what, every time dependence of the sample elongation on warmup consists of two contributions of different signs as depicted in Fig. 5, which indicates that there are two competing mechanisms, one of which is due to phonons while the other is presumably related to the dopant.

When studying thermal expansion of fullerite, in which octahedral voids were filled with heavy rare gas atoms like Ar or Xe, it was found [11,12] that the low-temperature expansion coefficients, if measured during warmup, are much lower than those measured during cooldown. In other words, a clearly detectable hysteresis has been documented as depicted in Fig. 6. The following explanation has been suggested. The low-temperature hysteretic phenomena were treated as manifestations of a first-order phase transition between several orientational glass states, which phenomenon is known as "polyamorphism". Transitions between those orientational-glass states are polyamorphic transformations, which can be of first order. However, it should be mentioned here that x-ray diffraction experiments [13] actually on the same sample revealed an incomparably larger hysteresis in the temperature dependence of the lattice parameter, spanning a range from 150 K across the phase transition point at 260 K to about

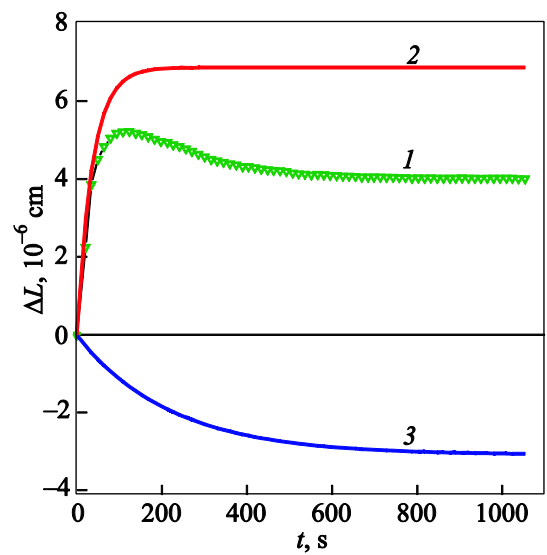

*Fig. 5.* Time dependence of the elongation $\Delta L$ of a $C_{60}$–$Xe_{0.3}$ sample upon heat pulse input, showing two thermal expansion contributions of opposite signs: experimental data (*1*); positive (*2*) and negative (*3*) contributions.

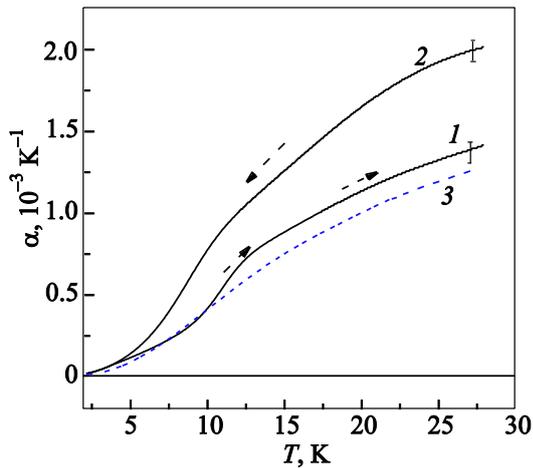

*Fig. 6.* Temperature dependence of the thermal expansion coefficient in the following experiments: warmup of $C_{60}$–$Xe_{0.3}$ (solid curve *1*); cooldown of $C_{60}$–$Xe_{0.3}$ (solid curve *2*). Dash curve *3* is for pure $C_{60}$.

300 K. On the other hand, since the temperatures are low enough to exclude thermal activation, the next assumption was that the rotations, which take place between orientational glass states, should have tunnel nature, which was also substantiated by negative expansivities. The relevant theory was developed by Bakai [8,11,14]. Later, Loktev and coworkers within a 2D model [15,16] showed that small-angle tunnel rotations are feasible in irregular areas, like dislocations or grain boundaries. An important consideration in the above approach was the following. A sufficiently large dopant particle expands the lattice, thereby lowering the energy barriers, which hamper molecular rotations. As a consequence, tunnel rotations become more likely bringing about negative expansion and making wider the temperature range of negative $\alpha$.

A different approach has been recently suggested [17] in which the negative expansion was ascribed entirely to dopants. The basis for this approach was two experimental facts: first, pure $C_{60}$ does not show negative expansion and, second, this effect has been observed [18] for more than ten dopant types, both molecular and atomic. In the case of a molecular dopant, the negative expansion could be easily explained by referring to Devonshire's model of a linear molecule in octahedral fields [19]. Yet, a similar effect was observed with rare gas dopants, even so heavy as xenon. This seemingly troublesome issue was explained [17] as being due to the fact that the first excited state of an atom in a three-dimensional well is a triplet, which is degenerate if the well is isotropic. But even in this case, the first triple-degenerate excited state can be represented as a set of appropriate combinations of three spherical harmonics, which corresponds to quantum rotation states. The potential energy $U$ of a Xe atom in an octahedral cage was evaluated using the known [20] Lennard-Jones potential between C and Xe atoms. The angular dependent part of the potential energy was truncated up to the rank-4 octahedral invariant: $U(r,\mathbf{w}) \simeq U_0(r) + \gamma U_4(r) I_4(\mathbf{w})$, where $r$ is the distance of the Xe atom from the void center; $\mathbf{w}$ designates angular variables; $U_0(r)$ is the isotropic part of the potential, which for Xe is very close to the harmonic one; $I_4(\mathbf{w})$ is the rank-4 octahedral invariant; $\gamma$ is the parameter, which for the particular case of Xe is known but was varied to obtain a general dependence. Account for the anisotropy of a real octahedral well results in a tunneling-related splitting (Fig. 7) of the energy spectrum resembling that of the Devonshire model. Thus, the negative contribution to the expansion coefficient can come from the transitions between levels of the manifold formed due to tunneling (the lower four curves) and, thus, are characterized by a negative Grüneisen parameter. On the other hand, transitions between different manifolds are characterized by positive Grüneisen parameters, providing an additional contribution to the positive expansivity as a function of temperature in the form of broad maxima.

Introduction of methane and deuteromethane molecules (for octahedral void occupancies about 50%) brought about a substantial (about 30%) decrease of thermal expansivities $\alpha$ (Fig. 8) compared to pure $C_{60}$, whereas rare gas doping led to higher $\alpha$ values as shown in Fig. 6. This finding was unexpected, since the van der Waals radii of both methane and deuteromethane compares with octahedral void size. It was known [21] that methane doping at low temperatures entails a measurable (about 0.12%) lattice expansion, which decreases the rotation barriers and thereby facilitates rotational tunneling. Therefore, the decrease of the thermal expansion in $C_{60}$ doped with the methanes can be related with tunnel molecular rotations, which in turn (see also Appendix) ensures a negative contribution to the thermal expansion. A stronger effect of $CH_4$ can be explained by its more pronounced quantum nature compared to the heavier isotope with its larger moment of inertia.

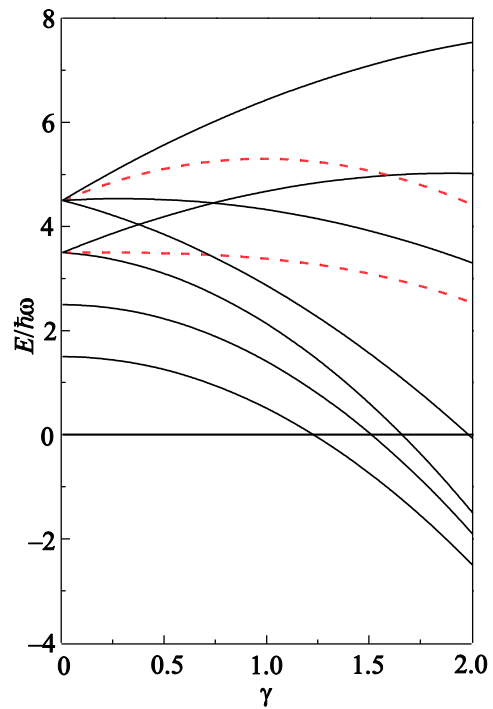

*Fig. 7.* Four lowest energy levels as a function of perturbation energy parameter $\gamma$, the parameter at the octahedral invariant. The dash curves are absent in Devonshire's theory.

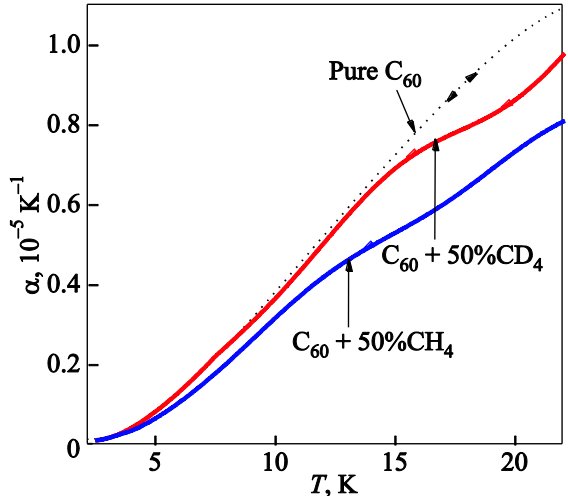

*Fig. 8.* "Cooldown" temperature dependence of the linear thermal expansion coefficient of fullerite $C_{60}$ doped with methane and deuteromethane.

Under warmup within the range 4 to 5.5 K both systems $(CH_4)_{0.5}$–$C_{60}$ and $(CD_4)_{0.5}$–$C_{60}$ showed peaks in the temperature dependence of $\alpha$ (Fig. 9). These features have been tentatively explained [22] as being due to a polymorphic orientational phase transformation point present within this temperature range. Evidence for this conclusion comes from sharp peaks in the temperature dependence of the characteristic times of the positive contribution to $\alpha$. Such peaks were observed (Fig. 10) for dopants characterized by a strong noncentral interaction with $C_{60}$ molecules.

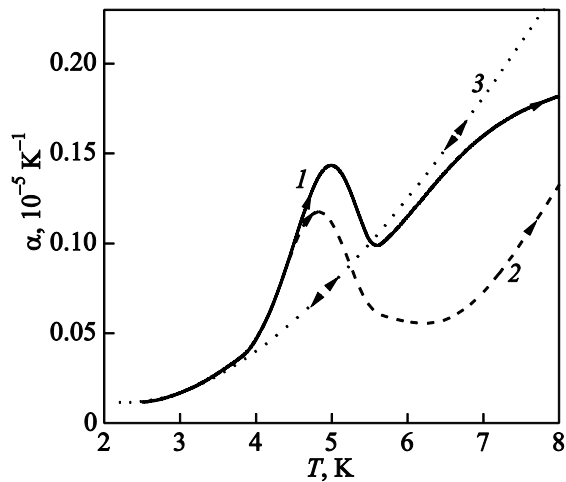

*Fig. 9.* Part of the "warmup" temperature dependence of the thermal expansion coefficient of fullerite $C_{60}$ doped with methane (curve *1*) and deuteromethane (curve *2*). Curve *3* is for pure $C_{60}$.

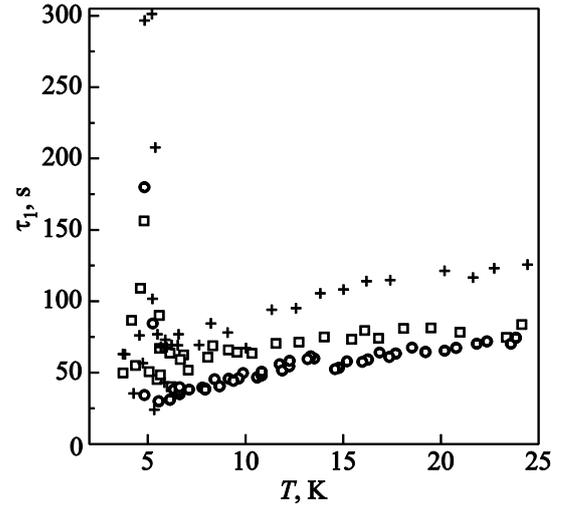

*Fig. 10.* Temperature dependence of the characteristic times for the positive contribution to thermal expansivity [22–24] in $C_{60}$ doped with CO and the methanes: $(CO)_{0.9}$–$C_{60}$ (circles); $(CH_4)_{0.5}$–$C_{60}$ (squares); $(CD_4)_{0.5}$–$C_{60}$ (crosses).

## 4. Diffusion of light-mass particles in $C_{60}$

Previous investigations of absorption rates of various gases in $C_{60}$ have been carried out at sufficiently high temperatures (from room to 300 °C), for otherwise the process would have been too slow and the resulting saturation levels, too low, at least from the technologic standpoint. Accurate measurements which can result in reasonably precise diffusion rates are long to measure, especially if the experiment is performed at low pressures to avoid overstuffing problems. It was established [25] that He infusion at room temperature into a $C_{60}$ powder with an average grain size of about 9.2 μm is a two-stage process as shown in Fig. 11. Knowing the grain size and deducing the diffusion time from a fit to the relevant diffusion theory curve, the diffusion coefficient was evaluated to be $D \simeq 7.5 \cdot 10^{-14}$ cm$^2$/s for the process of octa-void filling. Filling of tetrahedral voids occurs much slower during characteristic times which are an order of magnitude longer than during octa-filling.

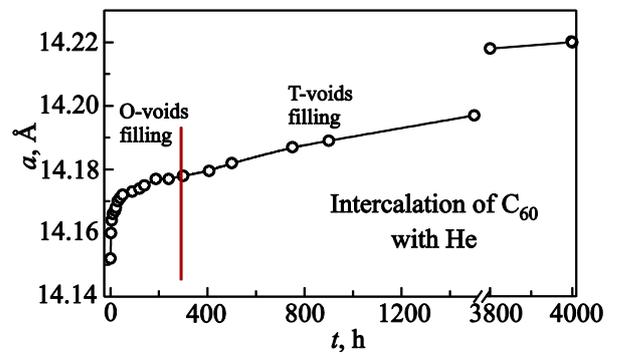

*Fig. 11.* The $C_{60}$ lattice parameter variation in time [28]. The brown vertical line demarcates the regions with prevailing O-void (left) and T-void (right) filling.

The process of He diffusion in fullerite $C_{60}$ over the wide temperature range from 2 to 300 K was studied [26] using the pressure monitoring technique mentioned in Sec. 2. The quantity measured was the pressure $P$ in the work cell as a function of time $t$ since the beginning of saturation or degassing. An example of experimental runs at three temperatures is shown in Fig. 12. The total amount of the He gas introduced before saturation was reasonably small, the helium to carbon atomic ratio not exceeding 0.1. It should be noted that the $\tau$ values obtained for degassing and saturation did not differ significantly. The pressure vs. time curves, like those shown in Fig. 12, were fitted to a sum of two exponentials with two characteristic times $\tau$ to be determined from fitting. These two sets $\tau(T)$ were treated as describing the kinetics of filling octahedral and tetrahedral voids with He atoms. Taking the typical grain size $r$ to be 1 μm, it was easy [26] to evaluate the diffusion coefficients from Einstein's relation $D = r^2/6\tau$ for both helium isotopes as depicted in Fig. 13.

It is worthwhile to compare the room-temperature diffusion coefficients derived from pressure monitoring (see Fig. 13) $D = 8 \cdot 10^{-12}$ cm$^2$/s and from x-ray diffraction [28] $D = (2.8 \pm 0.8) \cdot 10^{-14}$ cm$^2$/s, i.e., the former to latter $D$ value ratio is 285 while the corresponding grain size ratio squared used in estimates is 85, i.e., we come to an almost 3-fold difference, which is quite acceptable for diffusivity evaluations.

Since the general picture resembles very much corresponding curves for the quantum diffusion coefficient of $^3$He impurities in solid $^4$He in the regime of ultimately low $^3$He fractions [27,28] (see also Fig. 14), we recapitulate the main points of this phenomenon and corresponding considerations. At temperatures close to the melting point the diffusion is of purely thermally activated nature. If the $^3$He fraction is not low enough (as $X_1$ in Fig. 14), the deformation-related interaction between impurity atoms is suffi-ciently strong to render $D$ temperature independent below a certain temperature. In this regime, $D \propto 1/x$ at low temperatures [29]. In strongly diluted helium isotope solid mixtures, the interaction between impurities is sufficiently weak, which allows phonons to come into play (concentration $X_2$ and $X_3$ in Fig. 14), and here $D \propto T^{-9}$. The phonon contribution to the scattering cross-section diminishes steeply as the temperature goes down and at a sufficiently low $T$ the main role goes to the inter-impurity interactions. In this region, again, $D \propto 1/x$.

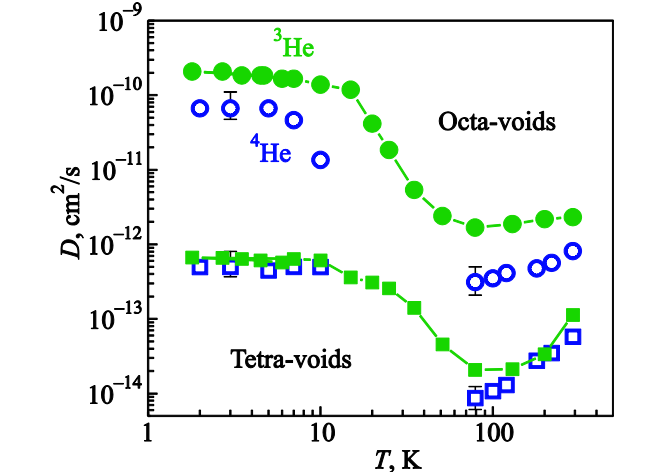

*Fig. 13.* Diffusion coefficients for the octahedral and tetrahedral voids for two helium isotopes. The circles are for octahedral voids (empty symbols for $^4$He, solid symbols for $^3$He), the squares are for tetrahedral voids (empty symbols for $^4$He; solid symbols for $^3$He).

Applying similar reasoning to what is depicted in Fig. 13, one can surmise that there is a clear analogy between the two phenomena under discussion. At higher temperatures (above roughly 100 K) the diffusion of He isotopes in $C_{60}$ is controlled by thermal activation. Below approximately 100 K the diffusion coefficient goes up with decreasing temperature. It should be remarked here that this turning point is very close to the temperature $T_g$ where the orientational glass state completes its formation.

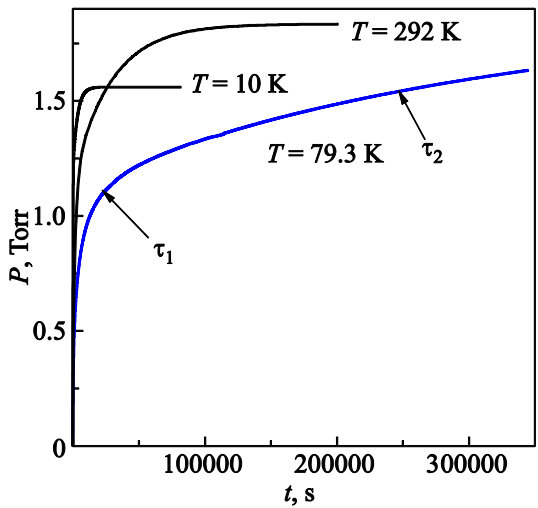

*Fig. 12.* Time dependence of the He gas pressure in the sample chamber during degassing for three typical temperatures.

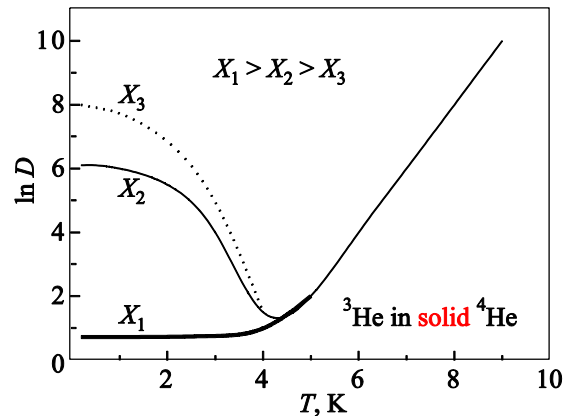

*Fig. 14.* Quantum diffusion of $^3$He in solid $^4$He for a range of $^3$He fractions.

Below the "turning point" near $T_g$ the diffusion coefficient increases quite fast as the temperature goes down, which is very possibly due to the diminishing role of phonons with decreasing temperature. As will be argued below, the fact that $T_g$ is a "turning point" is not fortuitous.

Although the energetics of the He atom in the lattice voids of $C_{60}$ was calculated in sufficient detail [30] in the atom-atom approximation with the relevant constant borrowed from data on He scattering on the graphite surface [31], no consistent theory is available so far.

There is clear evidence (see Fig. 15) of a quantum character of the diffusion of other light-mass particles such as hydrogen molecules and neon atoms [32]. These impurities fill only octahedral voids. Just like with the helium isotopes, increased diffusivities have been obtained for these dopants at temperatures below the orientational vitrification point $T_g$. By contrast, the diffusion coefficient of "classical" heavier particles like argon [33] goes down with decreasing temperature over the entire temperature range of experiment (58 to 290 K), which is evidence that the diffusion is a thermally activated process. Yet, below $T_g$ an essential decrease (from 410 K at higher temperatures to about 45 K at lower temperatures) of the activation energy occurs. This is another confirmation of the cardinal influence of abrupt changes in the rotational dynamics of $C_{60}$ molecules on the diffusion of dopant particles in fullerite.

Now concerning the role of the orientational glassification point $T_g$ in dopant transport over the lattice voids. In the case of a coherent (tunneling) process in a molecular crystal with internal degrees of freedom, two types of disorder, static and dynamic, control the situation. Static disorder can be due to any deviation from the ideal periodicity that shapes the dispersion law of diffusing particles as well as due to other tunneling particles, because, as scatterers, they are practically immobile. Dynamic disorder stems from time variation of the surrounding of tunneling particles. Usually, the main source of time-dependent variations are all kinds of phonons. However, in the case of $C_{60}$ there are unusual variations associated with molecular rotations, which results in specific effects inherent only in $C_{60}$. Thus, the sharp changes in the rotational dynamics across the phase transition at 260 K brings about a sharp response in the electrical conductivity [34]. In addition, when the temperature grows up across the orientational glassification point $T_g$ and molecular rotations are getting "defrozen", the photoluminescence emission starts to fade because rotational motion breaks up coherent transport of triplet excitions moving to traps [35]. Summing up, one may surmise that the turnover of the diffusion coefficient at $T_g$ as shown in Fig. 13 occurs because below that point the dynamic disorder stemming from molecular jump-like rotations ceases to exist and only phonons remain in force, their contribution diminishing fast with decreasing temperatures. The temperature-independent part of the $D(T)$ dependence at lowermost temperatures can be explained either by remaining static disorder of the orientational glass state or by the interaction between diffusing He atoms.

## 5. Radial expansion of carbon nanotubes

Among neat carbon nanotubes [36,37] it is the single-walled species that attract most attention of researches because they are least rigid and, as such, promise to support novel quantum properties in these hybrids of one- and two-dimensional systems.

As explained in Sec. 2, the cylinder-shaped compacted sample, depicted in Fig. 4, ensures to a high degree of reliability that the quantity to be measured is the total of the radial expansion of the nanotubes compressed in the sample. According to Schelling and Keblinski [38], owing to the fact that a nanotube, as a wrapped-up graphite plane, has a specific low-frequency phonon dispersion law, which must entail a negative Grüneisen parameter. The dilatometric measurements on a non-doped NT sample corroborated [39] those predictions: below 6 K down to 2 K the thermal expansion $\alpha$ turned out to be negative (black curve in Fig. 16). Saturation of bundles of single-walled carbon nanotube with various sufficiently heavy particles like Xe, $O_2$, $N_2$, etc. showed [40] that $\alpha$ retained its negativity but the region of negative values shrank towards lower $T$, as demonstrated in Fig. 16. The explanation is that the Xe atoms physisorbed on the outer surfaces of SWNT bundles suppress the bending phonon modes which are responsible for negative $\alpha$ values. Even doping NT bundles with hydrogen fits the picture.

When helium isotopes were used as dopants, the temperature dependence of the thermal expansion coefficient changed drastically. In Fig. 17 we show the radial expansivities of the nanotube ensemble in bundles doped with helium isotopes in comparison with the effect caused by xenon and molecular hydrogen; the He to C atomic ratio upon completion of the saturation process was about 9.4%. The effect of $^4$He doping proved [41] to be appreciable, to say the least: the lowermost negative α value in doped $C_{60}$ exceeded by more than an order of magnitude the value for the pure NT sample. Yet, the effect of the $^3$He doping [42] was absolutely miraculous, the minimum α value was roughly one hundred (!) times larger than for doping with the heavier helium isotope.

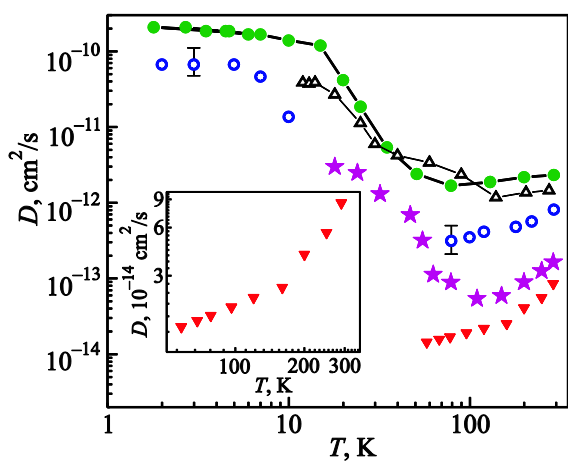

*Fig. 15.* Temperature dependence of the diffusion coefficients over octahedral voids of Ar (red downward triangles) [33], $^4$He (blue empty circles [26]), $^3$He (green solid circles [26]), $H_2$ (empty upright triangles [32]), and Ne (magenta stars [32]) in double logarithmic scale. Inset: data for argon doping within for a shorter $T$ range, showing a sharp kink in the $D(T)$ dependence.

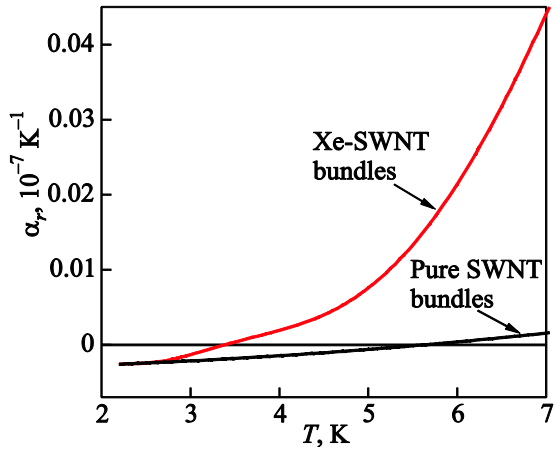

*Fig. 16.* Radial thermal expansion coefficients of pure and Xe-doped SWNT bundles at low temperatures.

Considering the overall wealth of the experimental findings we conclude that the most likely (and which, if not) reason behind the phenomena documented in expansivity measurements on the helium-doped sample is some quantum-mechanical mechanism, which is still unknown. Even plausible, however radical, suggestion are lacking. To get insight into the problem, the nature of helium atom dynamics on the surface of the specific single-walled nanotube, for which the interaction energy of a He atom with its surface was known [43], has been investigated theoretically [44]. It was shown that helium isotopes tunnel over the nanotube surface as a light-bound quasi-particle within a band 10 K (for $^4$He) or 14 K (for $^3$He) wide.

### Conclusions

In this short overview of recent findings in experiments on the saturation kinetics of the novel carbon nanomaterials, $C_{60}$ fullerite and single-walled nanotube bundles, we have shown that there is a wealth of new phenomena which could not be treated otherwise but as based on tunneling. In particular, below approximately 100 K the diffusion coefficients of the helium isotopes increase with decreasing temperature to become almost temperature independent roughly below 10 K, the entire situation resembling the behavior of the diffusion coefficient of $^3$He in solid $^3$He. Bundles of single-walled carbon nanotubes, when doped with $^4$He and, especially, $^3$He demonstrate negative radial expansion, which is also a sure signature of some tunnel mechanisms. The authors hope that the numerous aspects of the phenomena surveyed will serve a stimulus for theorists to develop models and approaches for a better understanding of the underlying physics.

### Appendix A: Grüneisen parameter

When the contribution of the several subsystems can be treated as independent, the thermal expansion coefficient can be represented in the form

$$\beta = \frac{k\chi_T}{V}\left[\gamma_{ph}C_{ph} + \gamma_{tun}C_{tun}\right], \quad (A.1)$$

where $\chi_T$ is the isothermal compressibility; $C_{ph}$ and $C_{tun}$ are the heat capacities due to phonons and tunneling states, respectively; $\gamma_{ph}$ and $\gamma_{tun}$ are the Grüneisen parameters for phonons and tunneling states, which are defined as

$$\gamma = \frac{\partial \ln \Delta E}{\partial \ln V}. \quad (A.2)$$

Here $\Delta E$ is the characteristic energy level difference for the corresponding subsystem. Thus, in rare gas solids the attractive energy (proportional to $V^{-2}$) determines the phonon thermodynamics, which results in typical values of γ close to 2. The tunneling energy splitting $\Delta E$ can be represented [45] in the form

$$\Delta E \propto \exp\left(-\sqrt{U(V)/\varepsilon}\right), \quad (A.3)$$

where ε is an expression of dimension "energy". It can be easily shown that the Grüneisen parameter for tunneling states as defined in Eq. (A.3) is negative and can reach values as high as a few hundreds.

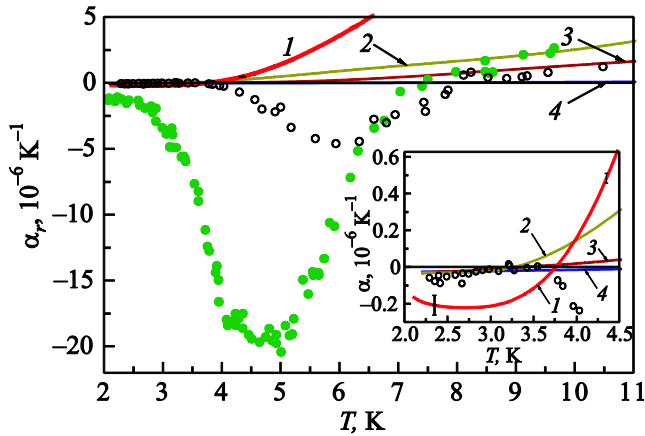

*Fig. 17.* Effect of doping on the thermal expansion of SWNT bundles: low-$T$ region. Curve *1* is for $^4$He; curve *2* is for $H_2$; curve *3* is for Xe; curve *4* is for pure NTs. Solid circles are α values for complete $^3$He doping; empty circles are for the sample after a ceratin part of dopant has been removed. Inset: a blowup of the 2 to 4.5 K range, showing in detail the negative α values of $^4$He-doped NTs.

Using high-precision experimental data on radial expansivities and heat capacities, obtained on the same compacted sample, the respective Grüneisen coefficient was estimated [46]. It turned out to be strongly temperature dependent, which is quite understandable, negative and equal to roughly −10 at 2 K, changing sign at approximately 6 K, and levelling fast off at a value of 4 at higher temperatures.

**Acknowledgment**

The authors express their gratitude to V.G. Manzhelii for his interest in this review and valuable remarks. Our thanks are also due to Razet Basnukayeva for helping with manuscript preparation.

1. H.W. Kroto, J.R. Heath, S.C. O'Brien, R.F. Curl, and R.E. Smalley, *Nature* **318**, 162 (1985).
2. K.S. Novoselov, A.K. Geim, S.V. Morozov, D. Jiang, Y. Zhang, S.V. Dubonos, I.V. Grigorieva, and A.A. Firsov, *Science* **306**, 666 (2004).
3. P.A. Heiney, J.E. Fischer, A.R. McGhie, W.J. Romarow, A.M. Denenstein, J.P. McCauley, Jr., A.B. Smith III, and D.E. Cox, *Phys. Rev. Lett.* **66**, 2911 (1991).
4. W.I.F. David, R.M. Ibberson, and T. Matsuo, *Proc. Roy. Soc.* (*London*) **442**, 129 (1993).
5. W.I. David, R.M. Ibberson, J.C. Matthewman, K. Prassides, T.J.S. Dennis, J.P. Hare, H.W. Kroto, R. Taylor, and D.R.M. Walton, *Nature* (*London*) **353**, 147 (1991).
6. W.I.F. David, R.M. Ibberson, T.J.S. Dennis, J.P. Hare, and K. Prassides, *Europhys. Lett.* **18**, 219 (1992).
7. A.M. Tolkachev, A.N. Aleksandrovskii, and V.I. Kuchnev, *Cryogenics* **15**, 547 (1975).
8. A.N. Aleksandrovskii, V.B. Esel'son, V.G. Manzhelii, B.G. Udovichenko, A.V. Soldatov, and B. Sundqvist, *Fiz. Nizk. Temp.* **23**, 1256 (1997) [*Low Temp. Phys.* **23**, 943 (1997)].
9. A.V. Dolbin, V.B. Esel'son, V.G. Gavrilko, V.G. Manzhelii, N.A. Vinnikov, S.N. Popov, and B. Sundqvist, *Fiz. Nizk. Temp.* **34**, 860 (2008) [*Low Temp. Phys.* **34**, 678 (2008)].
10. A.V. Dolbin, V.B. Esel'son, V.G. Gavrilko, V.G. Manzhelii, N.A. Vinnikov, S.N. Popov, N.I. Danilenko, and B. Sundqvist, *Fiz. Nizk. Temp.* **35**, 613 (2009) [*Low Temp. Phys.* **35**, 484 (2009)].
11. A.N. Aleksandrovskii, A.S. Bakai, D. Cassidy, A.V. Dolbin, V.B. Esel'son, G.E. Gadd, V.G. Gavrilko, V.G. Manzhelii, S. Moricca, and B. Sundqvist, *Fiz. Nizk. Temp.* **31**, 565 (2005) [*Low Temp. Phys.* **31**, 429 (2005)].
12. A.N. Aleksandrovskii, A.S. Bakai, A.V. Dolbin, G.E. Gadd, V.B. Esel'son, V.G. Gavrilko, V.G. Manzhelii, B. Sundqvist, and B.G. Udovidchenko, *Fiz. Nizk. Temp.* **29**, 432 (2003) [*Low Temp. Phys.* **29**, 324 (2003)].
13. A.I. Prokhvatilov, N.N. Galtsov, I.V. Legchenkova, M.A. Strzhemechny, D. Cassidy, G.E. Gadd, S. Moricca, B. Sundqvist, and N.A. Aksenova, *Fiz. Nizk. Temp.* **31**, 585 (2005) [*Low Temp. Phys.* **31**, 445 (2005)].
14. A.S. Bakai, *Fiz. Nizk. Temp.* **32**, 1143 (2006) [*Low Temp. Phys.* **32**, 868 (2006)].
15. V.M. Loktev, J.M. Khalack, and Yu.G. Pogorelov, *Fiz. Nizk. Temp.* **27**, 539 (2001) [*Low Temp. Phys.* **27**, 397 (2001)].
16. J.M. Khalack and V.M. Loktev, *Fiz. Nizk. Temp.* **29**, 577 (2003) [*Low Temp. Phys.* **29**, 429 (2003)].
17. M.A. Strzhemechny and I.V. Legchenkova, *Fiz. Nizk. Temp.* **36**, 470 (2010) [*Low Temp. Phys.* **36**, 370 (2010)].
18. A.V. Dolbin, *Quantum and Dimension Effects in Low-Temperature Thermal Expansion of Carbon Nanostructures*, DSc Dissertation, B. Verkin Institute for Low Temperature Physics and Engineering, Kharkov, Ukraine (2012).
19. A.F. Devonshire, *Proc. Roy. Soc. A* (*London*) **153**, 601 (1936).
20. G. Vidali and M.W. Cole, *Phys. Rev. B* **29**, 6736 (1984).
21. B. Morosin, R.A. Assink, R.G. Dunn, T.M. Massis, J.E. Schirber, and G.H. Kwei, *Phys. Rev. B* **56**, 13611 (1997).
22. A.V. Dolbin, N.A. Vinnikov, V.G. Gavrilko, V.B. Esel'son, V.G. Manzhelii, G.E. Gadd, S. Moricca, D. Cassidy, and B. Sundqvist, *Fiz. Nizk. Temp.* **35**, 299 (2009) [*Low Temp. Phys.* **35**, 226 (2009)].
23. A.V. Dolbin, V.B. Esel'son, V.G. Gavrilko, V.G. Manzhelii, N.A. Vinnikov, G.E. Gadd, S. Moricca, D. Cassidy, B. Sundqvist, *Fiz. Nizk. Temp.* **33**, 1401 (2007) [*Low Temp. Phys.* **33**, 1068 (2007)].
24. A.V. Dolbin, V.B. Esel'son, V.G. Gavrilko, V.G. Manzhelii, N.A. Vinnikov, G.E. Gadd, S. Moricca, D. Cassidy, and B. Sundqvist, *Fiz. Nizk. Temp.* **34**, 592 (2008) [*Low Temp. Phys.* **34**, 470 (2008)].
25. K.A. Yagotintsev, M.A. Strzhemechny, Yu.E. Stetsenko, I.V. Legchenkova, and A.I. Prokhvatilov, *Physica B* **381**, 224 (2006).
26. A.V. Dolbin, V.B. Esel'son, V.G. Gavrilko, V.G. Manzhelii, N.A. Vinnikov, and S.N. Popov, *Pisma ZhETF* **93**, 638 (2011) [*JETP Lett.* **93**, 577 (2011)].
27. V.F. Andreev and I.M. Lifshits, *Zh. Eksper. Theor. Fiz.* **56**, 2057 (1969).
28. V.N. Grigoriev, B.N. Esel'son, and V.A. Mikheev, *Fiz. Nizk. Temp.* **1**, 5 (1975) [*Sov. J. Low Temp. Phys.* **1**, 3 (1975)].
29. V.N. Grigoriev, B.N. Esel'son, V.A. Mikheev, V.A. Slyusarev, M.A. Strzhemechny, and Yu.E. Shul'man, *J. Low Temp. Phys.* **13**, 65 (1969).
30. R.S. Sinkovits and S. Sen, *Phys. Rev. B* **51**, 13841 (1995).
31. G. Vidali and M.W. Cole, *Phys. Rev. B* **29**, 6736 (1984).
32. A.V. Dolbin, V.B. Esel'son, V.G. Gavrilko, V.G. Manzhelii, N.A. Vinnikov, and S.N. Popov, *Fiz. Nizk. Temp.* **38**, 1216 (2012) [*Low Temp. Phys.* **38**, 962 (2012)].
33. A.V. Dolbin, V.B. Esel'son, V.G. Gavrilko, V.G. Manzhelii, N.A. Vinnikov, and R.M. Basnukaeva, *Fiz. Nizk. Temp.* **38** (2013) (to be published).
34. M.A. Strzhmechny and E.A. Katz, *Fullerenes* **12**, 281 (2004).
35. P.V. Zinoviev, V.N. Zoryansky, N.B. Silaeva, Yu.E. Stetsenko, M.A. Strzhmechny, and K.A. Yagotintsev, *Fiz. Nizk. Temp.* **38**, 923 (2012) [*Low Temp. Phys.* **38**, 732 (2012)].
36. S. Iijima, *Nature* **56**, 354 (1991).
37. S. Iijima and T. Ichihashi, *Nature* **363**, 603 (1993).
38. P.K. Schelling and P. Keblinski, *Phys. Rev. B* **68**, 035425 (2003).

39. A.V. Dolbin, V.B. Esel,son, V.G. Gavrilko, V.G. Manzhelii, N.A. Vinnikov, S.N. Popov, and B. Sundqvist, *Fiz. Nizk. Temp.* **34**, 860 (2008) [*Low Temp. Phys*. **34**, 678 (2008)].
40. A.V. Dolbin, V.B. Esel'son, V.G. Gavrilko, V.G. Manzhelii, N.A. Vinnikov, S.N. Popov, N.I. Danilenko, and B. Sundqvist, *Fiz. Nizk. Temp.* **35**, 613 (2009) [*Low Temp. Phys*. **35**, 484 (2009)].
41. A.V. Dolbin, V.B. Esel'son, V.G. Gavrilko, V.G. Manzhelii, N.A. Vinnikov, S.N. Popov, and B. Sundqvist, *Fiz. Nizk. Temp.* **36**, 797 (2010) [*Low Temp. Phys*. **36**, 635 (2010)].
42. A.V. Dolbin, V.B. Esel'son, V.G. Gavrilko, V.G. Manzhelii, N.A. Vinnikov, S.N. Popov, and B. Sundqvist, submitted to *Fiz. Nizk. Temp.*
43. L. Firlej and B. Kuchta, *Colloids and Surfaces A: Physicochem. Eng. Aspects* **241**, 149 (2004).
44. M.A. Strzhemechny and I.V. Legchenkova, *Fiz. Nizk. Temp.* **37**, 688 (2011) [*Low Temp. Phys*. **37**, 547 (2011)].
45. L.D. Landau and E.M. Lifshits, *Quantum Mechanics: Non-Relativistic Theory*, Butterworth-Heinemann, Oxford (1977).
46. M.I. Bagatskii, M.S. Barabashko, A.V. Dolbin, and V.V. Sumarokov, *Fiz. Nizk. Temp.* **38**, 667 (2012) [*Low Temp. Phys*. **38**, 523 (2012)].